\documentclass[prb,showpacs,twocolumn,superscriptaddress,aps,a4paper]{revtex4}
\usepackage{dcolumn}
\usepackage{amsmath}
\usepackage{graphicx}
\usepackage{latexsym}
\usepackage{amsfonts}
\usepackage{amssymb}
\DeclareGraphicsExtensions{.pdf,.gif,.jpg}

\newcommand{\be}{\begin{equation}}
\newcommand{\ee}{\end{equation}}
\newcommand{\beq}{\begin{eqnarray}}
\newcommand{\eeq}{\end{eqnarray}}

\begin{document}



\title{Time-dependent transport in graphene nanoribbons}





\author{Enrico Perfetto}
\affiliation{Dipartimento di Fisica, Universit\'a di Roma Tor
Vergata, Via della Ricerca Scientifica 1, I-00133 Rome, Italy}

\author{Gianluca Stefanucci}
\affiliation{Dipartimento di Fisica, Universit\'a di Roma Tor
Vergata, Via della Ricerca Scientifica 1, I-00133 Rome, Italy}
\affiliation{European Theoretical Spectroscopy Facility (ETSF)}

\author{Michele Cini}
\affiliation{Dipartimento di Fisica, Universit\'a di Roma Tor
Vergata, Via della Ricerca Scientifica 1, I-00133 Rome, Italy}
\affiliation{Consorzio Nazionale Interuniversitario per le Scienze
Fisiche della Materia, Unit\'a Tor Vergata, Via della Ricerca
Scientifica 1, 00133 Rome, Italy}




\begin{abstract}
We theoretically investigate the time-dependent ballistic
transport in metallic graphene nanoribbons after the sudden
switch-on of a bias voltage $V$. The ribbon is divided in three
different regions, namely two semi-infinite graphenic leads and a
central part of length $L$, across which the bias drops linearly
and where the current is calculated. We show that during the early
transient time the system behaves like a graphene bulk under the
influence of a uniform electric field $E=V/L$. In the undoped
system the current does not grow linearly in time but remarkably
reaches a temporary plateau with dc conductivity $\sigma_{1}=\pi
e^{2}/2h$, which coincides with the minimal conductivity of
two-dimensional graphene. After a time of order $L/v_{F}$ ($v_{F}$
being the Fermi velocity) the current departs from the first
plateau and saturates at its final steady state value with
conductivity $\sigma_{2}=2e^{2}/h$ typical of metallic nanoribbons
of finite width.

\end{abstract}

\maketitle


The recent isolation of single layers of carbon atoms\cite{graph1}
has attracted growing attention in the transport properties of
graphene-based devices. In these systems unconventional phenomena
like the half-integer quantum Hall effect\cite{qhe} and the Klein
tunneling\cite{klein} have been observed. Such peculiar behavior
stems from the relativistic character of the electrons in the
carbon honeycomb lattice. Closed to Dirac point the charge
carriers behave as two-dimensional (2D) massless Dirac
fermions\cite{review} and have very high mobility\cite{morozov}.
This fact has stimulated theoretical and experimental
investigations into graphenic ultrafast devices\cite{thz0} like
field effect transistors\cite{fet}, \textit{p-n} junction diodes
and THz detectors\cite{thz1,thz2}. In these systems it is crucial
to have full control of the electronic response after the sudden
switch-on of an external perturbation, and the study of the
real-time dynamics is becoming increasingly important. The
investigation of the transient response also is of fundamental
interest. One of the most debated aspects of the transport
properties of graphene is the minimum conductivity
$\sigma_{\mathrm{min}}$ at the Dirac point.
From the theoretical point of view the problem arises from the
fact that the value of $\sigma_{\mathrm{min}}$ is sensitive to the
order in which certain limits (zero disorder and zero frequency)
are taken\cite{ziegler}, thus producing different values around
the quantum $e^{2}/h$\cite{ziegler,peres,varlamov,beenakker}. Very
recently Lewkowicz and Rosenten overcame this ambiguity employing
a time-dependent approach in which the dc conductivity is
calculated by solving the quench dynamics of 2D Dirac excitations
after the sudden switching of a constant electric field.
Interestingly their approach does not suffer from the use of any
regularization related to the Kubo or Landauer formalism and
yields $\sigma_{\mathrm{min}}=\pi e^{2}/2h$. Despite the large
effort devoted to the study of the transport properties of
graphenic systems, a genuine real-time analysis which treats on
equal footing transients effects and the long-time response of
graphene nanoribbons in contact with semi-infinite reservoirs is
still missing.

\begin{figure}[h]
\includegraphics[height=6.1cm ]{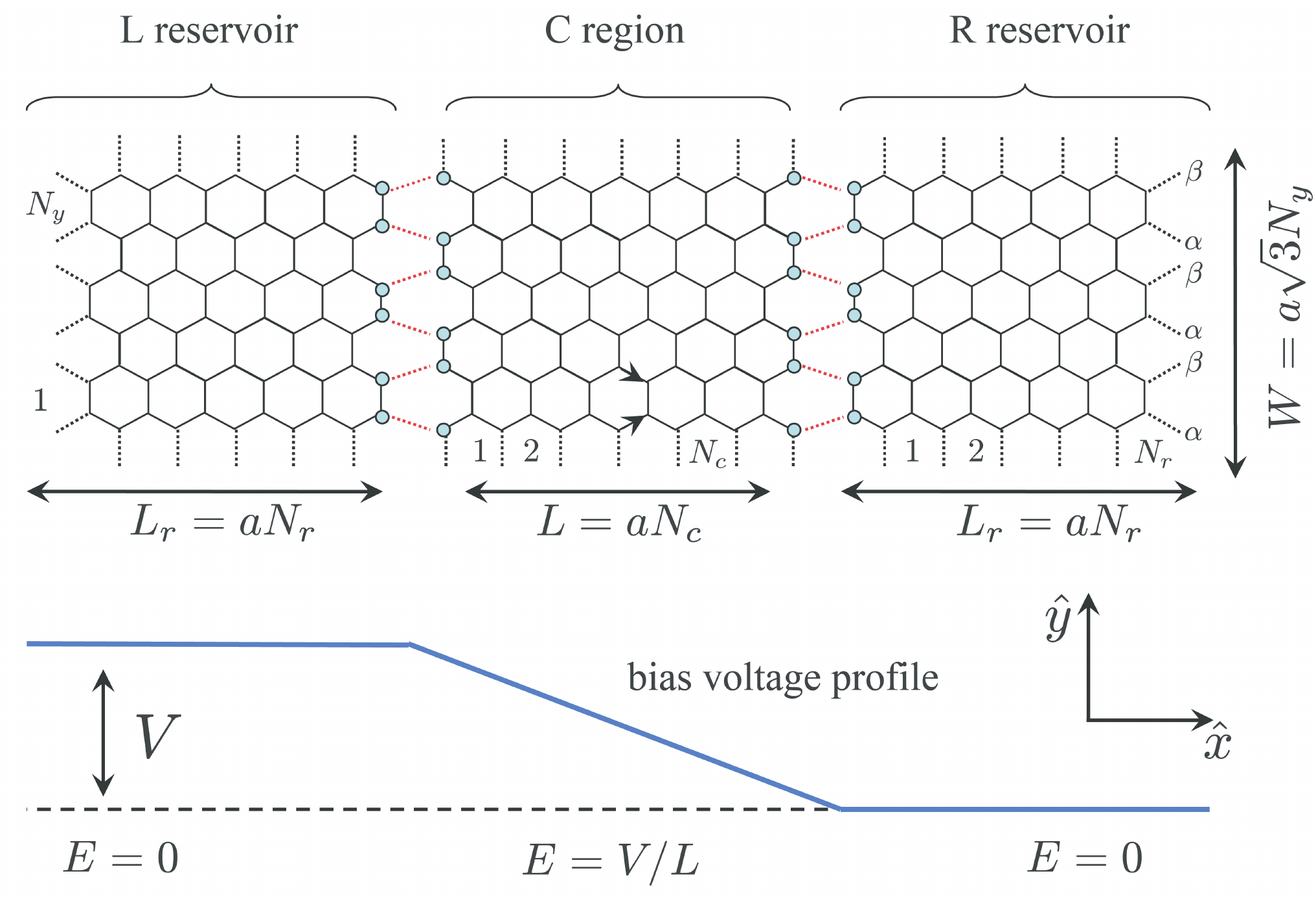}
\caption{Schematic representation of the system. The nanoribbon
has semi-infinite L a nd R graphenic reservoirs and periodic
boundary conditions are imposed along the $y$ direction. The bias
voltage profile with a linear drop inside the C region is also
shown.}
\label{fig1}
\end{figure}

In this Letter we study the time-dependent transport properties of
undoped graphene nanoribbons with finite width and virtually
infinite length after the sudden switch-on of an external bias
voltage. The geometry that we consider is sketched in
Fig.\ref{fig1}. The nanoribbon is divided in three regions, namely
a left (L) and a right (R) semi-infinite graphenic reservoirs and
a central (C) region of length $L=aN_{c}$, where $N_{c}$ is the
number of cells along the longitudinal $x$ direction and $a=2.46$
\AA $\,$ is the graphene lattice constant. The width of the ribbon
is $W=a\sqrt{3}N_{y}$, where $N_{y}$ is the number of cells along
the transverse $y$ direction, in which periodic boundary
conditions are imposed\cite{blanter,katnelson,guinea}.
The three regions are linked via transparent interfaces, in such a
way that in equilibrium the system is translationally invariant
along the $x$ direction, see Fig.\ref{fig1}.

Once the system is driven out of equilibrium, the quench dynamics
involves high energy excitations and the Dirac-cone approximation,
which is valid only at energies lower than 1 eV, is inaccurate.
For this reason we adopt a tight-binding description of the
system, with Hamiltonian given by
\begin{equation}
H=H_{0}+U(t)=v\sum_{\langle i,j \rangle }c^{\dagger}_{i}c_{j}+
\theta(t)\sum_{i}V_{i}c^{\dagger}_{i}c_{i} \, ,
\label{ham}
\end{equation}
where the spin index has been omitted and $v=2.7$ eV is the
hopping integral of graphene. The first sum runs over all the
pairs of nearest neighbor sites of the ribbon honeycomb lattice
and $c^{(\dagger)}_{i}$ is the annihilation (creation) operator of
a $\pi$ electron on site $i$. Here we use the collective index
$i=\{p, i_{y},i_{x} \}$ to identify a site in the nanoribbon such
that $p=\alpha,\beta$ indicates the two inequivalent longitudinal
zig-zag chains, $i_{y}$ denotes the cell in the $y$ direction, and
$i_{x}$ is the position in the $x$ direction. $H_{0}$ describes
the translationally invariant equilibrium system, while $U(t)$ is
the bias perturbation with (non self-consistent) voltage
profile\cite{lineardrop,lineardrop1,lineardrop2} given by the
function $V_{i}$
\begin{equation}
V_{i}=\left\{
\begin{array}{ll}
 V/2 & \quad i \in \mathrm{L}  \\
 V/2-E \, i_{x} &  \quad i \in \mathrm{C}\\
 -V/2 & \quad i \in \mathrm{R}
\end{array}
 \right. \,,
 \label{profile}
\end{equation}
where $V$ is the total applied voltage and $i_{x} \in (0,L)$ is
the $x$-coordinate of the site $i$ in the C region, which is
subject to the uniform electric field $E=V/L$ (see
Fig.\ref{fig1}). The above modelling of the bias profile could
find an approximate realization, e.g.,  in a planar junction in
which the L and R regions of the nanoribbon are on top of metallic
electrodes.

The time-dependent total current $I(t)$ flowing across the
interface in the middle of the C region is written as
\begin{equation}
I(t)
=2\sum_{i_{y}=1}^{N_{y}}\sum_{p=\alpha,\beta}I^{(p)}_{i_{y}}(t) \,
,
\end{equation}
where the factor 2 accounts for the spin degeneracy and
$I^{(p)}_{i_{y}}$ is the current flowing across the
$p=\alpha,\beta$ chain in the $i_{y}$-th cell, in the middle of
the C region (see Fig.\ref{fig1}).

Since periodic boundary conditions are imposed along the $y$
direction, the current $I^{(p)}_{i_{y}}$ does not depend on the
cell and chain indices $i_{y}$ and $p$, and hence
$I(t)=4N_{y}\bar{I}(t)$, where we have defined $\bar{I} \equiv
I^{(\alpha)}_{i_{y}}=I^{(\beta)}_{i_{y}}$ for any $i_{y}$.
\begin{figure}[h]
\includegraphics[height=3.5cm ]{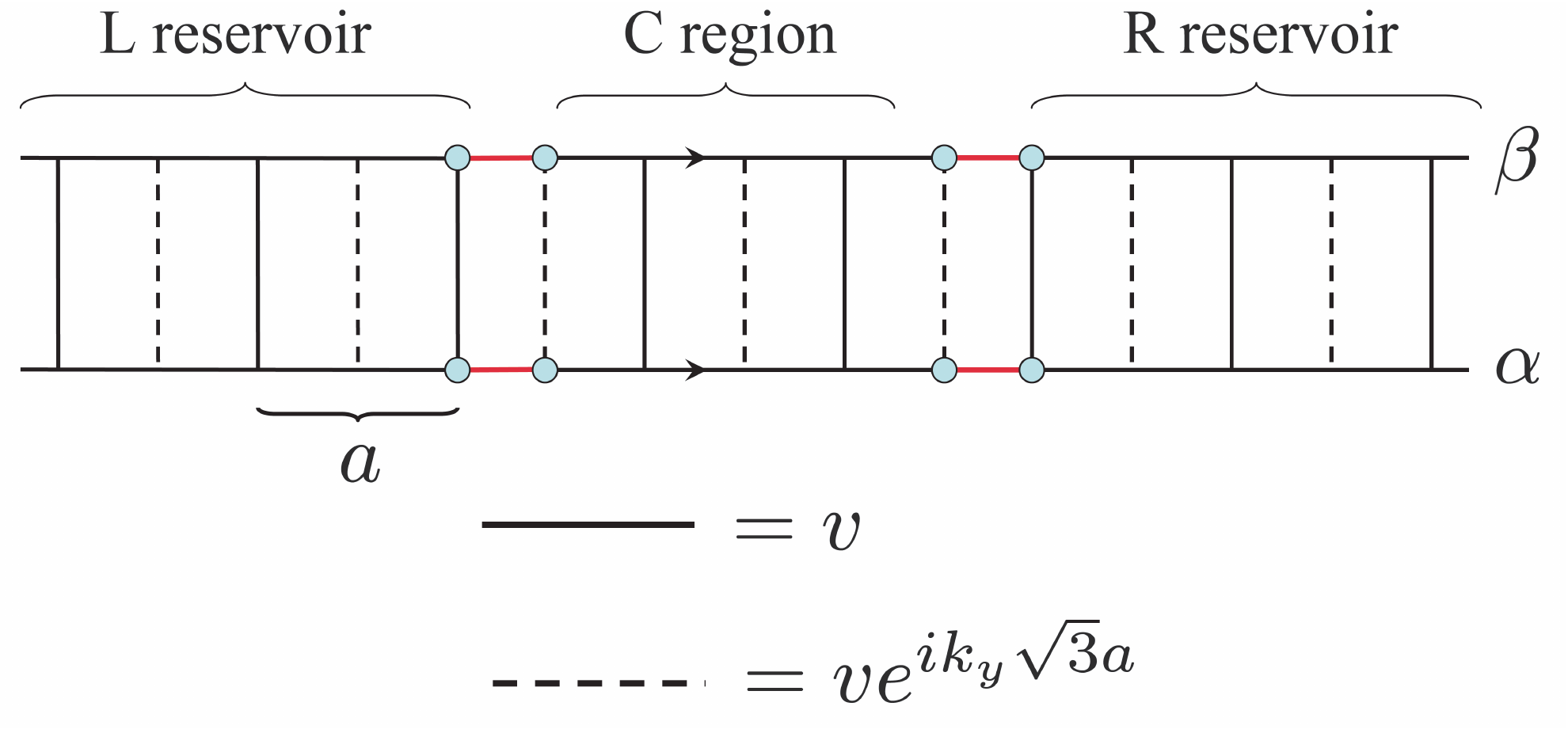}
\caption{Ladder model for fixed transverse momentum $k_{y}$. The
bias profile is the same as in Fig.\ref{fig1}.}
\label{fig1a}
\end{figure}
Since the transverse momentum $k_{y}=2\pi n/\sqrt{3}aN_{y}$ (with
$n=0,\dots N_{y}-1$) is conserved the current $\bar{I}$ can be
written as
\begin{equation}
\bar{I}(t)=\frac{1}{N_{y}}\sum_{k_{y}}\bar{I}_{k_{y}}(t) \,.
\label{currsum}
\end{equation}
It is seen that the total current of the original $N_{y}$-wide
problem can be calculated by summing the currents
$\bar{I}_{k_{y}}$ coming from $N_{y}$ independent 1-wide ladder
problems\cite{blanter}  with staggered $k_{y}$-dependent
transverse hopping (see Fig.\ref{fig1a}), with the same bias
profile as in Eq.(\ref{profile}).

In order to calculate $\bar{I}_{k_{y}}$ it is convenient to
introduce the $y$-Fourier transform of the original electron
operators $c_{\{p,k_{y},i_{x} \}}=(N_{y})^{-1/2}
\sum_{i_{y}}e^{ik_{y}i_{y}\sqrt{3}a}c_{\{p,i_{y},i_{x} \}}$ in
terms of which the current $\bar{I}_{k_{y}}$ can be cast as
\begin{equation}
\bar{I}_{k_{y}}(t)=\frac {2\, e\, v}{\hbar}\mathrm{Re} \left[
G^{<}_{\{p,k_{y},\frac{L}{2}\};\{p,k_{y},\frac{L}{2}+\frac{a}{2}\}}(t,t)
\right] \,,
\end{equation}
where $G^{<}$ is the lesser Keldysh Green's function
\begin{equation}
G^{<}_{\{p,k_{y},i_{x}\};\{r,q_{y},j_{x}\}}(t_{1},t_{2})=i\langle
c^{\dagger}_{\{p,k_{y},i_{x} \}} (t_{1}) c_{\{r,q_{y},j_{x}\}}
(t_{2})\rangle \,.
\end{equation}
The time-evolution of the Green's function is evaluated according
to
\begin{eqnarray}
G^{<}_{\{p,k_{y},i_{x}\};\{p,k_{y},j_{x}\}}(t,t)= \nonumber \\
i \left[ e^{-iH t} f(H_{0}) e^{iH t}
\right]_{\{p,k_{y},i_{x}\};\{p,k_{y},j_{x}\}} \, ,
\label{evol}
\end{eqnarray}
where we recall that there is no actual dependence on $p$ and
where $f$ is the Fermi distribution function. For each transverse
momentum $k_{y}$ the current $\bar{I}_{k_{y}}$ is numerically
calculated by computing the exact time evolution of the
corresponding ladder system in Fig.\ref{fig1a}, where we take the
reservoirs with a \textit{finite} length $L_{r}=a N_{r}$. This
approach allows us to reproduce the time evolution of the
infinite-leads system up to a time $T_{\mathrm{max}} \approx
2L_{r} /v_{F}$, where $v_{F}$ is the Fermi velocity\cite{perf}.
For $t > T_{\mathrm{max}}$ electrons have time to propagate till
the far boundary of the leads and back, yielding undesired finite
size effects in the calculated current. Accordingly we choose
$N_{r}$ such that $T_{\mathrm{max}}$ is much larger than the time
at which the steady-state is reached.
\begin{figure}[h]
\includegraphics[height=4.5cm ]{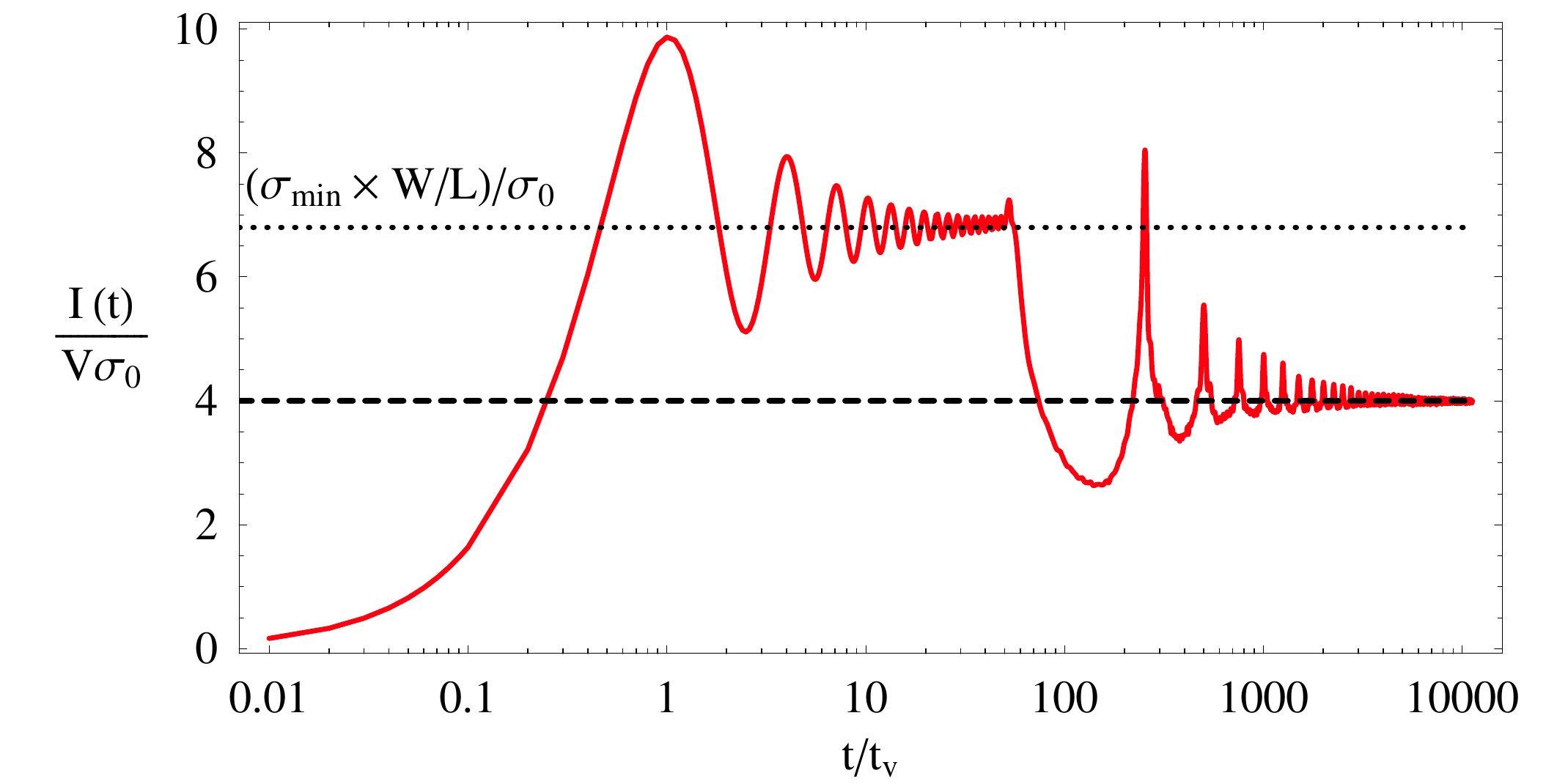}
\caption{Total time dependent current $I(t)$ with geometric
parameters $L\approx 25$ $n$m, $W \approx 106$ $n$m and applied
voltage $V=8\times10^{-4}$ Volt.
Current is in units of $V \sigma_{0}$, where $\sigma_{0}=e^{2}/h$
is the quantum of conductance and time in units of
$t_{v}=2.5\times 10^{-16}$ s. The dashed line represents the
long-time conductance $\sigma_{2}/\sigma_{0}=4$ given by the
Landauer formula\cite{onipko2}, while the dotted line represents
the combination $(\sigma_{\mathrm {min}} \times
W/L)/\sigma_{0}=\pi/2 \times W/L $.}
\label{fig2}
\end{figure}

For practical purposes we represent $H_{0}$ and $H(t)$ in a hybrid
basis, in which the Hamiltonians of the isolated (equilibrium and
biased) L and R reservoirs are diagonal in the set of states
derived analytically in Ref.\onlinecite{onipko}, while the
Hamiltonian describing the C region remains represented in the
basis $\{p,k_{y},i_{x}\}$. The enormous advantage of such choice
is clarified in the following. Since we are interested in the dc
conductance we set a small bias $V<<\hbar v_{F}/eW$. In this way
we ensure that at long times a linear regime is established in
which only the electrons right at the Dirac point (with $k_{y}=0$)
contribute to the total current. This agrees with the Landauer
formula, according to which only the states within the bias window
contribute at the steady-state. On the other hand during the
transient \textit{all} transverse modes are excited and the sum in
Eq.(\ref{currsum}) must be computed including the complete set of
$k_{y}$. Indeed we have checked numerically that in the limit
$t\rightarrow \infty $ all currents $\bar{I}_{k_{y}}$, except the
one with $k_{y}=0$, vanish. We have also observed that the damping
time of $\bar{I}_{k_{y}>0}(t)$ goes like $k_{y}^{-1}$ and the
calculation of the currents with small $k_{y}$ requires a very
long propagation before the zero-current steady-state is
approached. As a consequence very large values of $N_{r}$ are
needed, making the computation in principle too demanding. Such
numerical difficulty is, however, compensated by the fact that,
after a transient time of order $L/v_{F}$, for $k_{y}$ close to
the Dirac point only few low-energy states contribute to
$\bar{I}_{k_{y}}(t)$. Therefore for any given $k_{y}>0$ we
introduce an energy cutoff $\Lambda_{k_{y}}\approx 10 \hbar v_{F}
k_{y}$ in the reservoirs Hamiltonians and retain only the
lead-eigenstates with transverese momentum $k_{y}$ and energy in
the range $(-\Lambda_{k_{y}},\Lambda_{k_{y}})$. This allows us to
deal with very long leads ($N_{r}>>1$) and can be explicitly
implemented by using the analytic eigenstates of a rectangular
graphenic macromolecule\cite{onipko}. The C region (which is the
one where we calculate the current) is treated exactly within the
original full basis $\{p,k_{y},i_{x}\}$, but the overall
computational cost remains moderate.

\begin{figure}[h]
\includegraphics[height=4.5cm ]{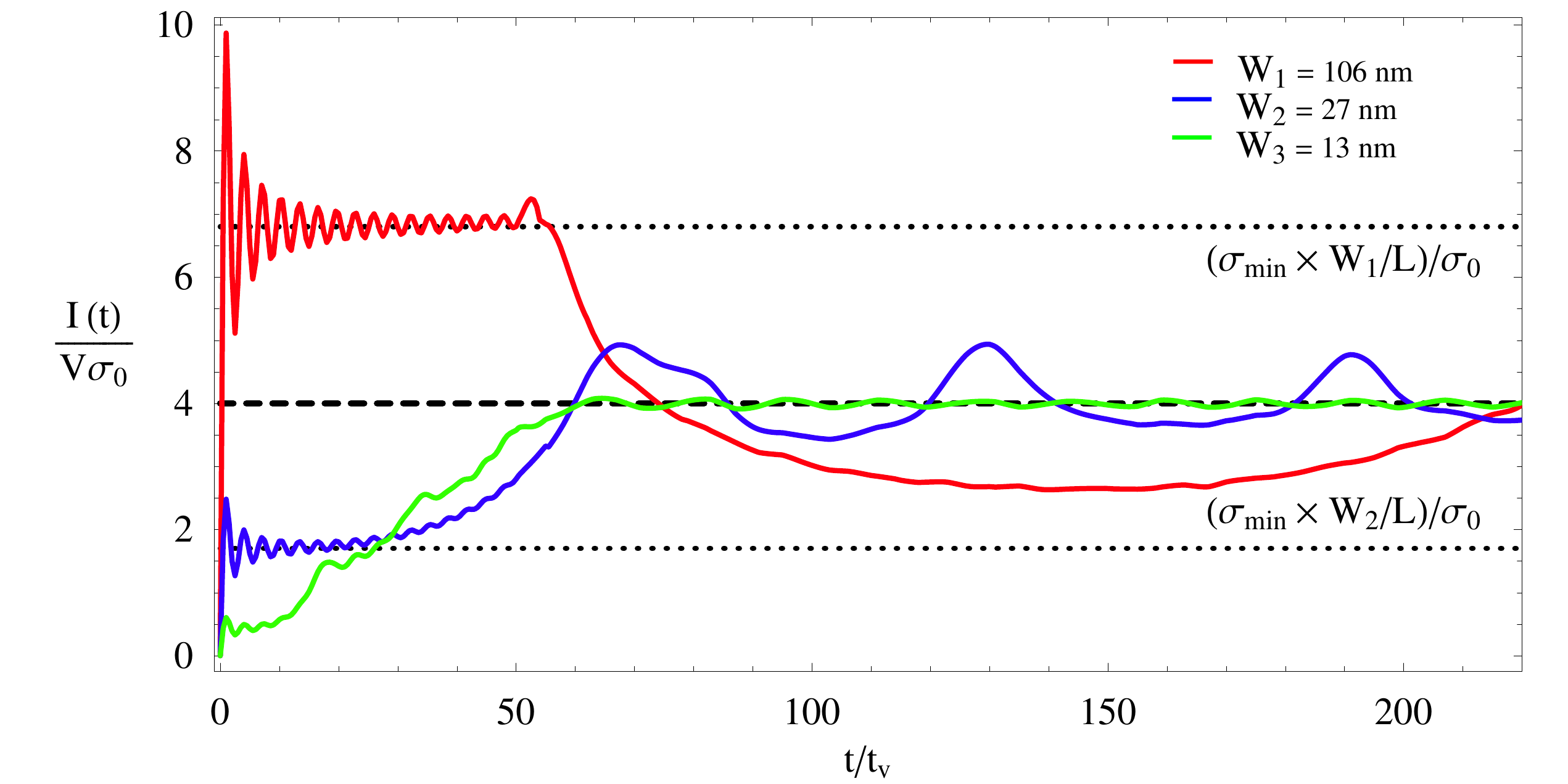}
\caption{Short time transient current $I(t)$ for different ribbon
widths $W_{1}=106$ $n$m, $W_{2}=27$ $n$m, $W_{3}=13$ $n$m.  The
rest of parameters, dotted/dashed lines and units are as in
Fig.\ref{fig2}.} \label{fig3}
\end{figure}

In Fig.\ref{fig2} we show the time-dependent current $I(t)$
calculated for a nanoribbon of central length $L\approx 25$ $n$m
($N_{c}=100$) and width $W \approx 106$ $n$m ($N_{y}=250$) with an
applied voltage $V=8\times10^{-4}$ Volt and zero temperature. The
very early transient regime ($t\lesssim \hbar/v \equiv t_{v}$)
depends on the details with which the electric field $E$ has been
switched-on (sudden in time in the present case). For $ t_{v}
\lesssim t \lesssim  L/v_{F}$, however the time evolution only
depends on the geometry of the device and on the potential
profile. Within this time domain, if the ribbon is wide enough,
the ballistic electron dynamics of the system probes only the bulk
properties of graphene, since the particles do not have time to
explore the reservoirs, where $E=0$. According to the Drude
picture, the ballistic transport in bulk materials subjected to
uniform electric fields produces a time-dependent conductance
given by
\begin{equation}
\sigma(t)=\frac{\gamma t}{1+(\omega t)^{2}} \, , \label{drude}
\end{equation}
where $\gamma$ is a constant proportional to the density of
states. In ordinary solids the dc conductance ($\omega \rightarrow
0$) increases linearly in time up to the breakdown of the
ballistic regime, in which one has to replace $t$ with the finite
scattering time $\tau$. However in pure graphene the transport is
ballistic up to very large length-scales of the order of micron
and $\sigma$ can apparently diverge, producing a Drude peak.
Nevertheless in undoped graphenic samples the density of states at
the Fermi level vanishes, thus making the product $\gamma t $
constant at long times. This subtle compensation is at the origin
of the finite minimal dc conductivity of 2D pure
graphene\cite{lew} and of the difficulties in constructing a
suitable propagation scheme in finite width nanoribbons. In
Fig.\ref{fig2} we see that $W \approx 100$ $n$m is enough to
observe such compensation. The current $I(t)$, instead of
increasing linearly in time, reaches a temporary plateau with
average current $I_{1}$ which lasts till $t \approx L/v_{F}
\approx 60 t_{v}$. On the contrary in an ordinary system (e.g. in
a 2D square lattice model) the current would be linear in time up
to $L/v_{F}$, producing the well known Drude peak in the bulk
limit $L,W \to \infty $ at fixed $E$. We would like to observe
that the plateau at $I_{1}$ is reached via transient oscillations
with frequency $2v /\hbar$\cite{lew}. Interestingly such frequency
is not displayed by any of the individual currents
$\bar{I}_{k_{y}}(t)$, but appears only as a cumulative effect
after the summation in Eq.(\ref{currsum}) is performed. Therefore
it is a a genuine bulk property and may be at the origin of the
resonant effect predicted to occur in optical response of graphene
right at $\omega=2 v/\hbar$\cite{acgraphene}.

\begin{figure}[h]
\includegraphics[height=4.5cm ]{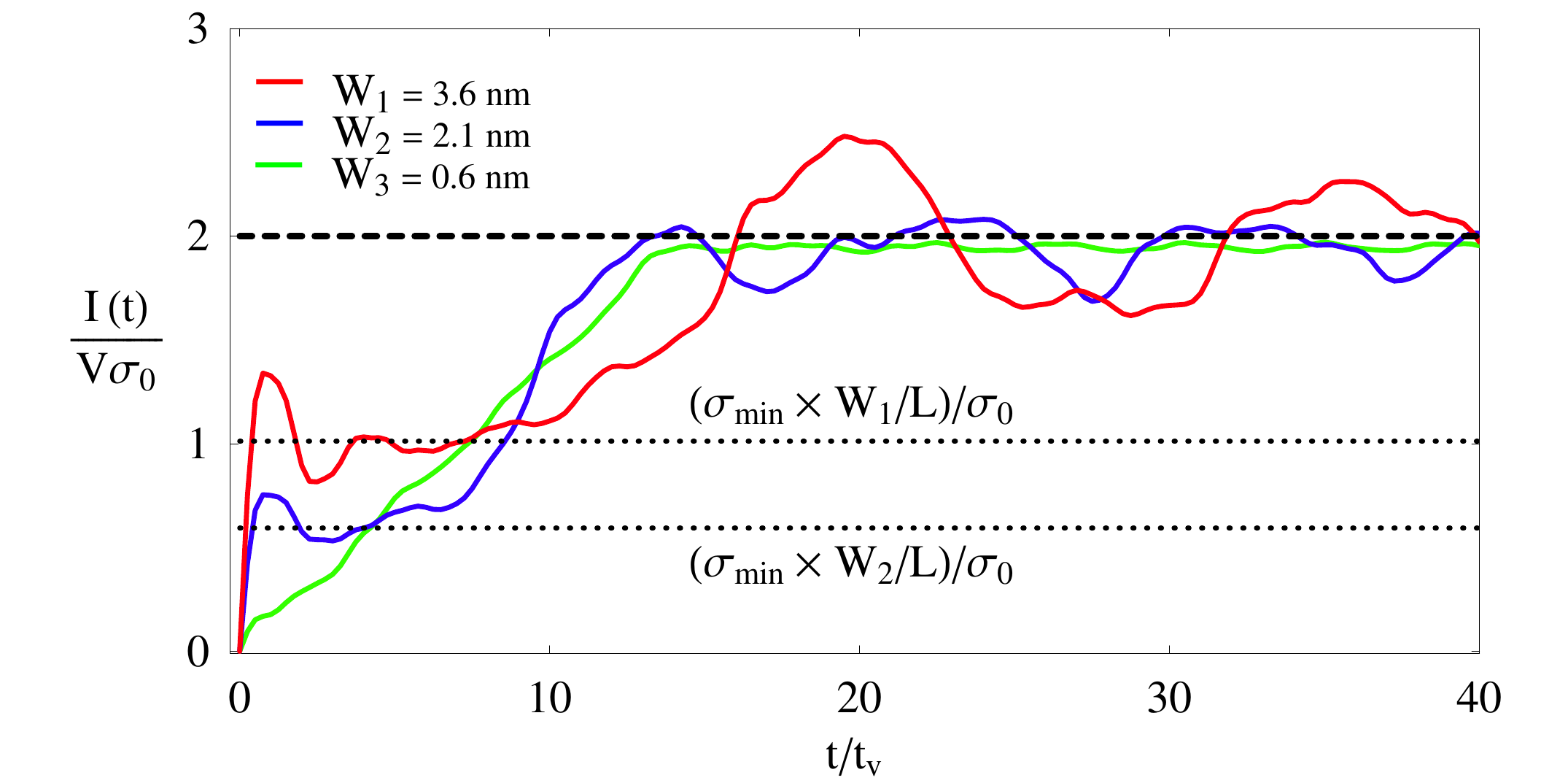}
\caption{Short time transient current $I(t)$ for three different
metallic armchair ribbons with open boundary conditions. The
ribbon parameters are $L=5.5$ $n$m, $W_{1}=3.6$ $n$m, $W_{2}=2.1$
$n$m, $W_{3}=0.6$ $n$m and the applied bias is $V=0.03$ Volt.
Units and dotted lines are as in Fig.\ref{fig2}. The dashed line
represents the long-time conductance $\sigma_{2}/\sigma_{0}=2$
given by the Landauer formula\cite{peres,onipko2}.}
\label{fig4}
\end{figure}

From the first plateau of $I(t)$ we can provide an independent
evaluation of the minimal conductivity of graphene. According to
its definition, the conductivity $\sigma_{1}$ of a bulk system
subjected to a small constant electric field $E$ is given by the
current density $J$ divided by $E$:
\begin{equation}
\sigma_{1}=\frac{J}{E}=\frac{I_{1} }{V }\frac{L}{W} \, .
\label{sigma1}
\end{equation}
By exploiting Eq.(\ref{sigma1}) it is seen that our data are
consistent with the value $\sigma_{1} = \pi e^{2}/2h \equiv
\sigma_{\mathrm{min}}$ with excellent precision, see
Fig.\ref{fig2}.

Further we have investigated the formation of the first plateau as
a function of the ribbon width $W$. In Fig.\ref{fig3} we see that
for narrow ribbons with $W \approx 10$ $n$m finite-size effects
produce a drastic deviation from the ideal graphene bulk. The
transient current does not show a temporary saturation, but grows
with an approximate linear envelope (up to a time $\sim L/v_{F}$),
in qualitative agreement with the Drude behavior given in
Eq.(\ref{drude}).

At times larger than $\sim L/v_{F}$ the electrons start exploring
the reservoirs, where the electric field is zero. At this point a
standard dephasing mechanism sets in and the current tends to its
true final steady-state. As discussed above, however, such value
is reached after a very slow damping process, in which the current
displays decaying oscillations with dominant frequency
$\bar{\omega} = 2 \pi v_{F} /W$  (see Fig.\ref{fig2}), $\hbar
\bar{\omega}$ being the energy spacing between the transverse
energy subbands of the ribbon. We have checked numerically that
the asymptotic value $I_{2}$ agrees well with the Landauer formula
and does not depend on $L$ and $W$, provided that $V <<\hbar
v_{F}/eW$. Thus the conductance of the device is simply extracted
as $\sigma_{2}=I_{2} /V$. Our numerical data provides
$\sigma_{2}=4e^{2}/h$ with high numerical accuracy. This value is
indeed twice the conductance of metallic nanoribbons, and this is
due to the chosen periodic boundary conditions\cite{onipko2}. Thus
we have calculated $I(t)$ also in the case of open boundary
conditions. Unfortunately, as discussed above, the lack of
translational invariance along the transverse direction makes the
computation much more demanding, and only system with small $L$
and $W$ can be studied within the present approach. In
Fig.\ref{fig4} we show $I(t)$ for three different metallic
armchair nanoribbons with open boundaries. It can be seen that
already for $W \approx 2-4 $ $n$m there is a tendency to form the
universal first plateau leading to $\sigma_{\mathrm{min}}$, while
for $W \lesssim 1 $ $n$m the current $I(t)$ grows linearly in time
until $t \approx L/v_{F}$. On the other hand at long times the
current tends clearly to the Landauer value, consistent with
$\sigma_{2}=2e^{2}/h$, independently on the aspect
ratio\cite{peres,onipko}.

In summary we pointed out the subtle difficulties in constructing
a reliable method to perform time-evolutions of finite width
graphene nanoribbons and proposed an efficient numerical scheme to
overcome them. We presented a real-time study of the transport
properties of these systems in contact with virtual semi-infinite
reservoirs in the linear regime. We have shown that for large
enough undoped samples the time-dependent current displays two
plateaus. From the first of these plateaus we can extract an
independent measure of the minimal conductivity $\pi e^{2}/2$ of
bulk graphene by resorting the aspect ratio $L/W$ of the device.
The second plateau corresponds to reaching the steady-state and is
independent of the geometry. Here the conductance is $2e^{2}/h$,
which coincides with the Landauer result for metallic nanoribbons.
To conclude we wish to point out that in presence of ac bias, the
time-dependent conductivity can be used to obtain the optical
conductivity $\sigma_{\mathrm{ac}}$ of graphene. It was shown
experimentally\cite{ac1} and explained theoretically\cite{ac2}
that $\sigma_{\mathrm{ac}}$ is almost $\omega$-independent and
equals $\sigma_{\mathrm{min}}$ with high accuracy over a wide
range of frequencies. Remarkably to extract the universal value of
$\sigma_{\mathrm{ac}}$ high frequency signals with $\omega \sim
1/t_{v}$ have been employed\cite{ac1}. Therefore we believe that a
real-time approach like to one presented here is needed to
enlighten the crossover from the dc case to ultrafast scenarios in
which the period of the ac signal is comparable with the intrinsic
hopping time of the bulk system.

%


\end{document}